\documentclass[cits]{PoS}
\usepackage[latin9]{inputenc}
\usepackage{array}
\usepackage{units}
\usepackage{amsmath}
\usepackage{amssymb}
\usepackage{graphicx}

\makeatletter

\providecommand{\tabularnewline}{\\}
\usepackage{slashed}

\title{Neutrinoless double $\beta$
  decay with small neutrino masses}

\ShortTitle{Large $0\nu\beta\beta$ with small $m_{\nu}$}

\author{F. del Águila$^a$,  A.  Aparici$^b$, S. Bhattacharya$^c$, \speaker{A. Santamaria}$^b$ and J. Wudka$^c$\\
\llap{$^a$}CAFPE and Departamento de Física Teórica y del Cosmos, Universidad de Granada, \\ 
E-18071 Granada, Spain\\
\llap{$^b$}Departament de Física Teòrica and IFIC, Universitat de València-CSIC,\\ 
Dr. Moliner 50, E-46100 Burjassot (València), Spain\\
\llap{$^c$}Department of Physics and Astronomy, University of California, Riverside CA 92521-0413, USA\\
E-mail: \email{faguila@ugr.es},  \email{Alberto.Aparici@uv.es},\email{subhaditya123@gmail.com},  \email{Arcadi.Santamaria@uv.es},  \email{jose.wudka@ucr.edu}
}

\abstract{If the only contribution to  neutrinoless double beta decay ($0\nu\beta\beta$) are neutrino masses its discovery may be very difficult, especially if neutrinos present a normal hierarchy spectrum. However, this is not the only possibility; $0\nu\beta\beta$  can in general produce electrons of either chirality, in contrast with the neutrino induced $0\nu\beta\beta$ which predicts two left-handed electrons. Using an effective Lagrangian approach we classify the lepton number violating (LNV) effective operators with two leptons of either chirality but no quarks, ordered according to the magnitude of their contribution to $0\nu\beta\beta$. We point out that, for each of the three chirality assignments, $e_{L}e_{L},e_{L}e_{R}$ and $e_{R}e_{R}$ , there is only one LNV operator of the corresponding type to lowest order, and these have dimensions 5, 7 and 9, respectively. Neutrino masses are always induced by these extra operators but can be delayed to one (dimension 7) or two loops (dimension 9). Under the assumption that $0\nu\beta\beta$ is dominated by the operators of dimension 7 or 9 we find that the scale of new physics should be relatively low ($\lesssim 30$~TeV). We also list the SM additions generating these operators upon integration of the heavy modes, and discuss simple realistic examples of renormalizable theories for each case. The phenomenology of a model giving rise to the dimension 9 operator has been analyzed with some detail: if $0\nu\beta\beta$  is going to be seen in the next round of experiments, the doubly charged scalars of the model could be seen at the LHC and lepton flavour violating (LFV) rates could be at the reach of foreseen experiments. Moreover neutrino masses, which arise at two loops,  display a very particular structure and are strongly constrained, in fact,  $\sin^{2}\theta_{13}\gtrsim0.008$, when $\mu\rightarrow eee$
is required to lie below its present experimental limit.}

\FullConference{Proceedings of the Corfu Summer Institute 2012 "School and Workshops on Elementary Particle Physics and Gravity"\\
September 8-27, 2012\\
Corfu, Greece}

\makeatother

\begin{document}
\global\long\def\vevof#1{\left\langle #1\right\rangle }

\section{Introduction}

The remarkable observation of neutrino oscillations (see \cite{Beringer:1900zz}
and \cite{Mohapatra:2005wg,GonzalezGarcia:2007ib} for recent reviews
and \cite{GonzalezGarcia:2012sz,Fogli:2012ua,Tortola:2012te,Machado:2011ar,Bergstrom:2012yi}
for fits including latest data on $\theta_{13}$) has provided the
first evidence for neutrino masses. On the other hand, the invisible
decay width of the Z-boson tells us that there are only three species
of light active neutrinos (lighter than about $45\,$GeV and interacting
with full gauge strength). If lepton number (LN) is conserved, neutrinos
are Dirac fields composed by the three active (left-handed) and three
sterile (right-handed) neutrinos. In that case, neutrinos are accommodated
in the Standard Model (SM) as the charged fermions, and their masses
and mixings, like in the quark sector, can be parametrized by the
three masses, $m_{1,2,3}$, three mixing angles, $\theta_{12},\,\theta_{23},\,\theta_{13}$,
and one CP-violating phase, $\delta$. If LN is not conserved and
there are no light sterile neutrinos the neutrino mass sector is completely
different from the quark sector but still can be parametrized by the
three masses, three mixing angles, the phase $\delta$ and two additional
phases, $\alpha_{1,2}$, which are characteristic of Majorana neutrinos.
Neutrino oscillation experiments only depend on mass differences,
the mixing angles and the LN conserving phase $\delta$, but not on
the Dirac/Majorana character of the neutrinos. Thus, present data
on neutrino oscillations allow us to determine rather precisely the
mass differences, $\Delta m_{21}^{2},\,|\Delta m_{31}^{2}|$, and
the mixing angles, %
\footnote{In the standard parametrization the sign of $\Delta m_{21}^{2}=m_{2}^{2}-m_{1}^{2}$
can be always chosen positive by convention. This is not true for
$\Delta m_{31}^{2}=m_{3}^{2}-m_{1}^{2}$ once the mixing angles are
taken to vary only in the first quadrant. The solution with $\Delta m_{31}^{2}>0$
is usually named normal hierarchy (NH) while $\Delta m_{31}^{2}<0$
is called inverted hierarchy (IH). %
} $\,\theta_{12},\,\theta_{23},\,\theta_{13}$, but little is known
on the phase, $\delta$, the sign of $\Delta m_{31}^{2}$ or the absolute
neutrino mass scale (characterized, for instance, by the lightest
neutrino mass). 

However, from the conceptual point of view, the main question is whether
neutrinos are Dirac or Majorana, for the symmetries they preserve
are different and require quite different descriptions in quantum
field theory. In fact, if there are no sterile neutrinos, active neutrinos
should necessarily be Majorana neutrinos, LN should not be conserved
and the SM should be extended with new particles in order to allow
for lepton number violation (LNV). 
\begin{figure}[h]
\begin{centering}
\includegraphics[width=0.55\textwidth]{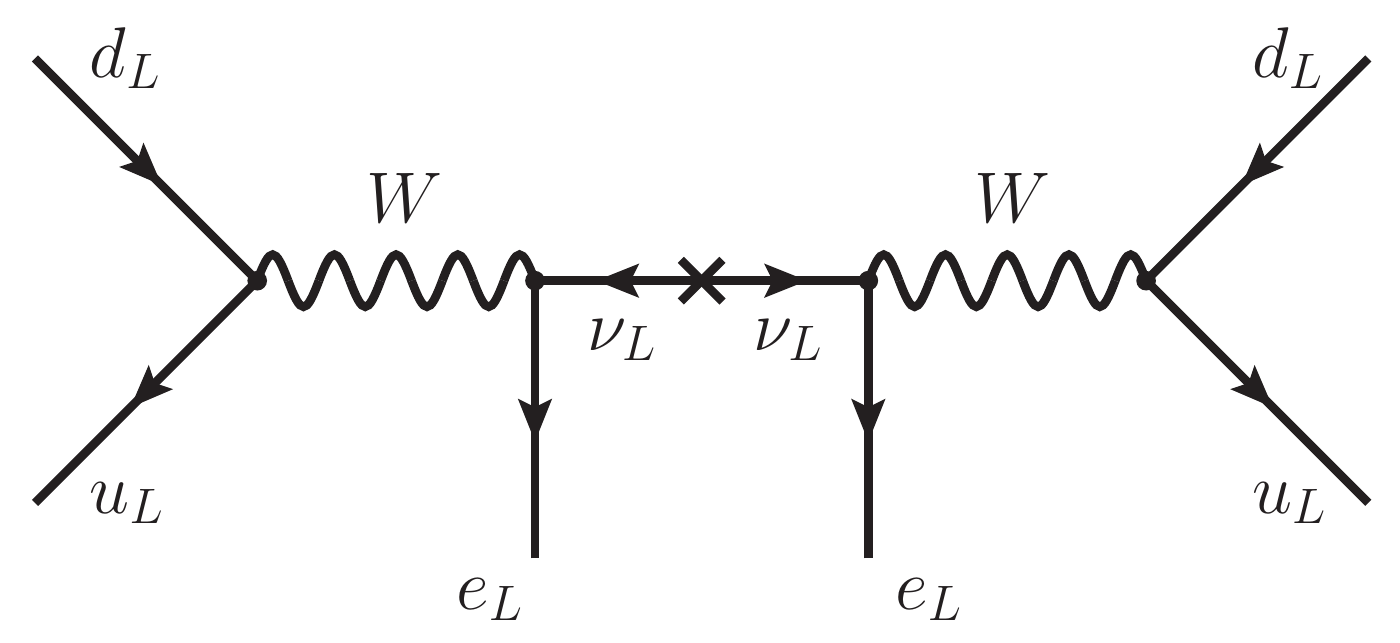}
\par\end{centering}

\caption{Majorana neutrino mass contribution to $0\nu\beta\beta$. \label{fig:numassTo0nbb}}
\end{figure}
Can one test if neutrino masses are of the Majorana type? It seems
very difficult since we have only observed neutrino masses in neutrino
flavour oscillations, which conserve total lepton number. LNV neutrino
transitions (for instance, a $\mu^{+}$ produced by a $\nu_{\mu}$
from $\mu^{-}$ decay) are in principle possible for Majorana neutrinos,
but, unless other sources of LNV are present, they are suppressed
by $m_{\nu}/E$ factors and do not enjoy an oscillatory behaviour
\cite{Pontecorvo:1957cp,Li:1981um,Schechter:1980gk,Bernabeu:1982vi,Langacker:1998pv}.
Alternatively, one can look for LNV decays. Since Majorana neutrinos
violate LN, searching for LNV we are indirectly testing the Majorana
character of the neutrinos. The best candidate is neutrinoless double
beta decay ($0\nu\beta\beta$) which is also suppressed by $m_{\nu}$
but is enhanced with respect to the LN conserving process (double
beta decay with emission of two neutrinos), which has been already
observed, by huge phase space factors. In Figure~\ref{fig:numassTo0nbb}
we display the standard Majorana neutrino contribution to $0\nu\beta\beta$.
Assuming no new physics (NP) beyond the SM and only three Majorana
neutrinos the present/future situation can be summarized by plotting
the allowed regions in the plane $\langle m_{\nu}\rangle-m_{\mathrm{MIN}}$,
where $\langle m_{\nu}\rangle$ is a combination of masses and mixings
relevant in~$0\nu\beta\beta$ and $m_{\mathrm{MIN}}$ is the lightest
neutrino mass. Under these assumptions, present experiments are not
able yet to probe neutrino masses. Planned experiments, however, will
be able to do it if neutrinos present an IH spectrum and/or $m_{\mathrm{MIN}}$
is above a few tens of meV. If $m_{\mathrm{MIN}}$ is below $10\:\mathrm{meV}$
and neutrino masses present a NH spectrum, it will be very difficult
to test LNV in $0\nu\beta\beta$. 

\begin{figure}
\begin{centering}
\includegraphics[clip,width=0.7\columnwidth]{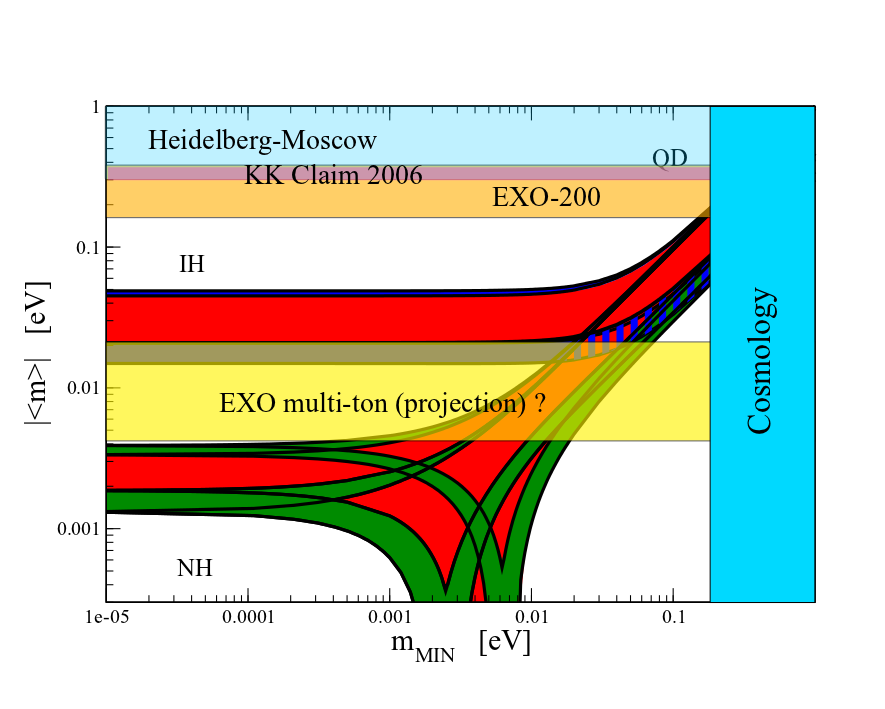}
\par\end{centering}

\caption{Prospects for testing $0\nu\beta\beta$ if induced only by Majorana
neutrino masses.\label{fig:Prospects-for-0nbb}}
\end{figure}
\begin{figure}
\begin{centering}
\includegraphics[width=0.45\textwidth]{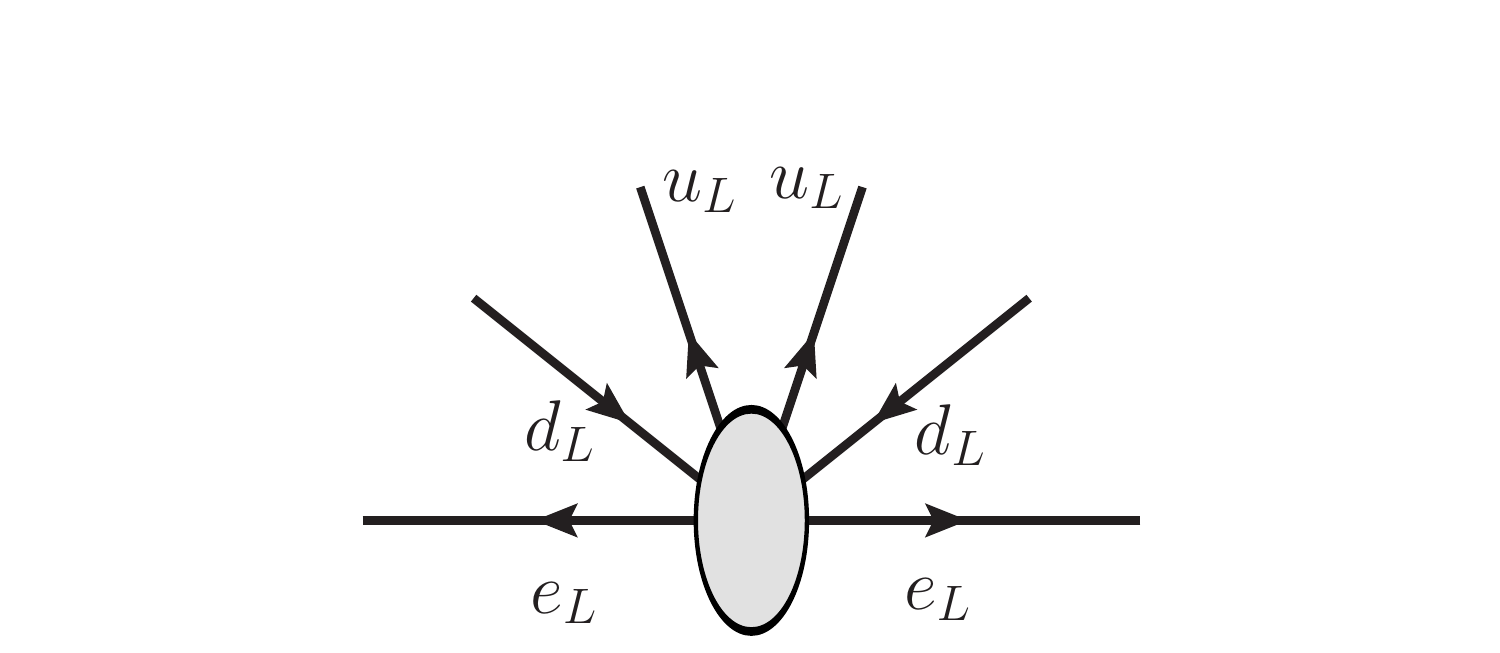}
~~~ \includegraphics[width=0.45\textwidth]{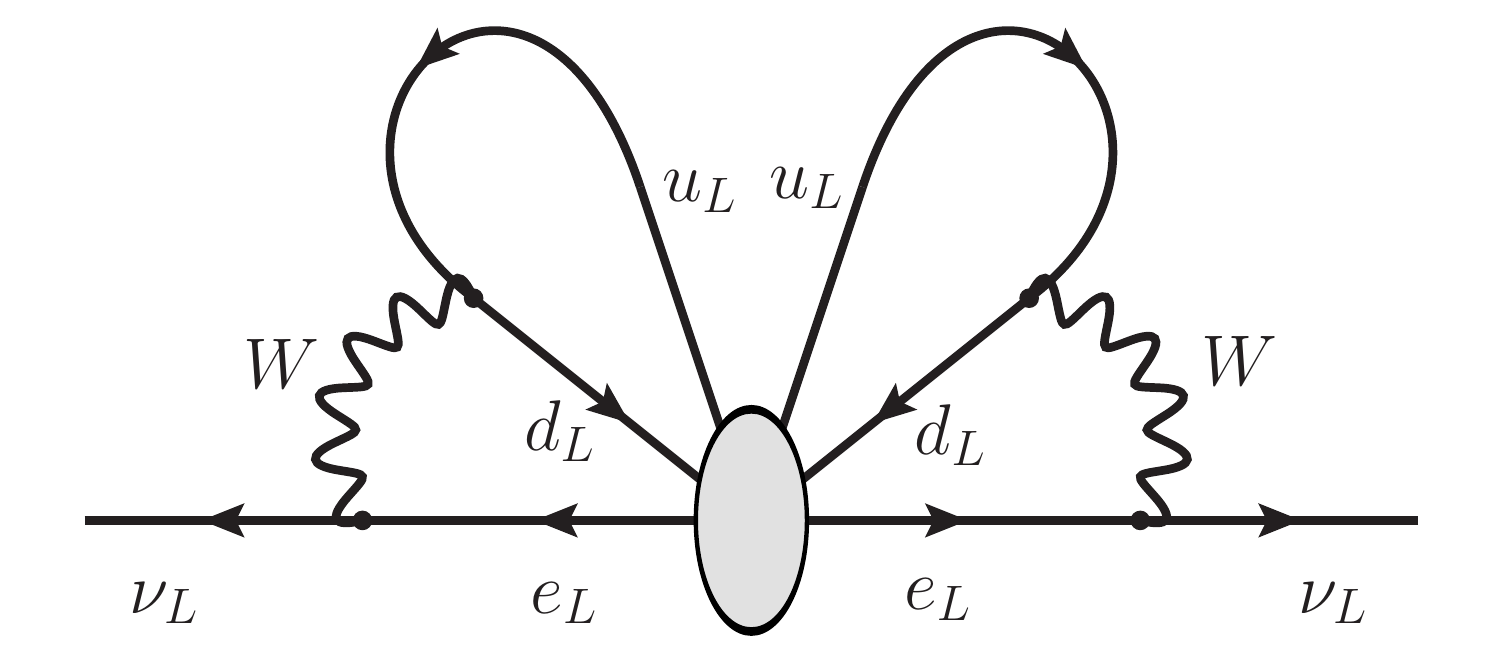}
\par\end{centering}

\caption{How $0\nu\beta\beta$ induces Majorana neutrino masses (at four loops).\label{fig:0nbbToMajorana} }
\end{figure}

On the other hand, the relation between Majorana neutrino masses and
$0\nu\beta\beta$ is not so direct in general, and there could be
new contributions to $0\nu\beta\beta$ which could render it observable
in planned experiments and perhaps even in upgrades of ongoing ones.
In fact, from the symmetry point of view, Majorana masses break any
charge carried by $\nu_{L}$ by two units, while $0\nu\beta\beta$
can proceed in different ways depending on the chirality of the emitted
electrons, $e_{L}e_{L}$, $e_{L}e_{R}$ or $e_{R}e_{R}$. Hence, this
decay will break any charge carried by $e_{L}$ and $e_{R}$ by $(2,0)$,
$(1,1)$ and $(0,2)$ units, respectively. Therefore, Majorana neutrino
masses and $0\nu\beta\beta$ could have, in principle, a quite different
origin. However, as in the SM $\nu_{L}$ and $e_{L}$ are in the same
multiplet, they should carry the same type of lepton number while
$e_{L}$ and $e_{R}$ are only linked by the electron mass. This just
tells that the connection between the different lepton numbers is
valid only up to SM gauge symmetry and/or chirality breaking effects.
Moreover, there is a general argument \cite{Schechter:1981bd} which
suggests that if $0\nu\beta\beta$ exists there should necessarily
be contributions to Majorana neutrino masses induced by the $0\nu\beta\beta$
interactions (see Figure~\ref{fig:0nbbToMajorana}). Nevertheless,
these contributions are generated at four loops and are expected to
be very small. 

In view of this discussion it is important to explore the possibility
of additional contributions to $0\nu\beta\beta$ and their connection
with neutrino masses, a scenario to be tested by existing and forthcoming
experiments. Since both, neutrino masses and $0\nu\beta\beta$, are
low energy processes, an effective Lagrangian approach \cite{Coleman:1969sm,Weinberg:1978kz,Weinberg:1980wa,Polchinski:1983gv,Georgi:1994qn,Wudka:1994ny}
is the proper starting point of any model-independent analysis. Several
papers~\cite{Babu:2001ex,Choi:2002bb,Engel:2003yr,deGouvea:2007xp}
have followed this approach, considering only effective interactions
not involving SM gauge bosons. Here we review a different class of
theories where NP does not couple directly to the quark sector%
\footnote{Operators involving leptons and quarks with no gauge bosons generate
neutrino masses at 1 to 4 loops~\cite{deGouvea:2007xp,Babu:2010vp,Duerr:2011zd,Babu:2011vb}
and may receive enhancements from top Yukawa couplings.%
}, so that the effective interactions involve only leptons and gauge
bosons (couplings to gauge bosons are generated whenever the NP is
not a SM gauge singlet). In this case it is possible to give a simple
classification of the effects that concern us in terms of only three
operators, each of which can be generated at tree level by different
types of NP \cite{delAguila:2012nu}. In the unitary gauge these operators
give the vertices $\nu_{L}\nu_{L},~We_{R}\nu_{L}$ and $WWe_{R}e_{R}$
and have dimension 5,7 and 9, respectively. This allows for three
scenarios wherein one of the operators is generated at tree level
and the others via loops. A simple example exhaustively considered
in the literature has tree-level neutrino masses generated via a high-scale
see-saw mechanism~\cite{Minkowski:1977sc,GellMann:1980vs,Yanagida:1979as,Mohapatra:1979ia},
with effective $We_{R}\nu_{L}$ and $WWe_{R}e_{R}$ vertices generated
radiatively. Here we concentrate on the complementary case where the
LNV operators contributing to $0\nu\beta\beta$ are generated at tree
level whereas neutrino masses are induced radiatively. It is also
very important to be able to build renormalizable models which could
give rise to the different operators in the effective Lagrangian at
low energies. As we will stress, these models are quite interesting
because some of the new particles may be accessible to LHC \cite{delAguila:2011gr,delAguila:2013yaa}. 

Thus, using the effective Lagrangian approach, in Section \ref{sec:New-Physics-Contributions}
we review the classification of the NP contributions to $0\nu\beta\beta$
and their connection with neutrino masses \cite{delAguila:2012nu}.
In Section \ref{sec:Renormalizable-Completions} we will discuss some
examples of renormalizable completions giving rise to the relevant
operators presented in Section \ref{sec:New-Physics-Contributions}
\cite{delAguila:2012nu,delAguila:2011gr}. Finally, Section~\ref{sec:Conclusions}
is devoted to our conclusions. Further details of the topics discussed
in this talk and a more complete list of references can be found in
these papers.

\section{NP Contributions to $0\nu\beta\beta$\label{sec:New-Physics-Contributions}}

The observation of $0\nu\beta\beta$ requires LNV, while SM interactions
conserve LN. Therefore, new interactions must be added to the SM in
order to allow for $0\nu\beta\beta$. We illustrate the different
possibilities diagrammatically in Figure~\ref{fig:0nbb-diagrams},
where only one LNV vertex is considered at a time. The new interactions
are represented by a big dot and the $e$ could be $e_{L}$ or $e_{R}$
depending on the particular vertex. Diagram $A$ is the usual contribution
provided by Majorana neutrino masses, diagrams $D$--$F$ require
new interactions involving quarks which will not be considered here
anymore. Therefore, we will concentrate on contributions of type $B$
and $C$ and in their relation with neutrino masses (contributions
of type $A$). Figure~\ref{fig:0nbb-diagrams} represents the low
energy form of the different interactions, but they must be obtained
preserving the SM gauge symmetries. For this we should use the effective
Lagrangian approach which, being very general, still involves some
assumptions:

\begin{figure}
\begin{centering}
\includegraphics[width=0.9\textwidth]{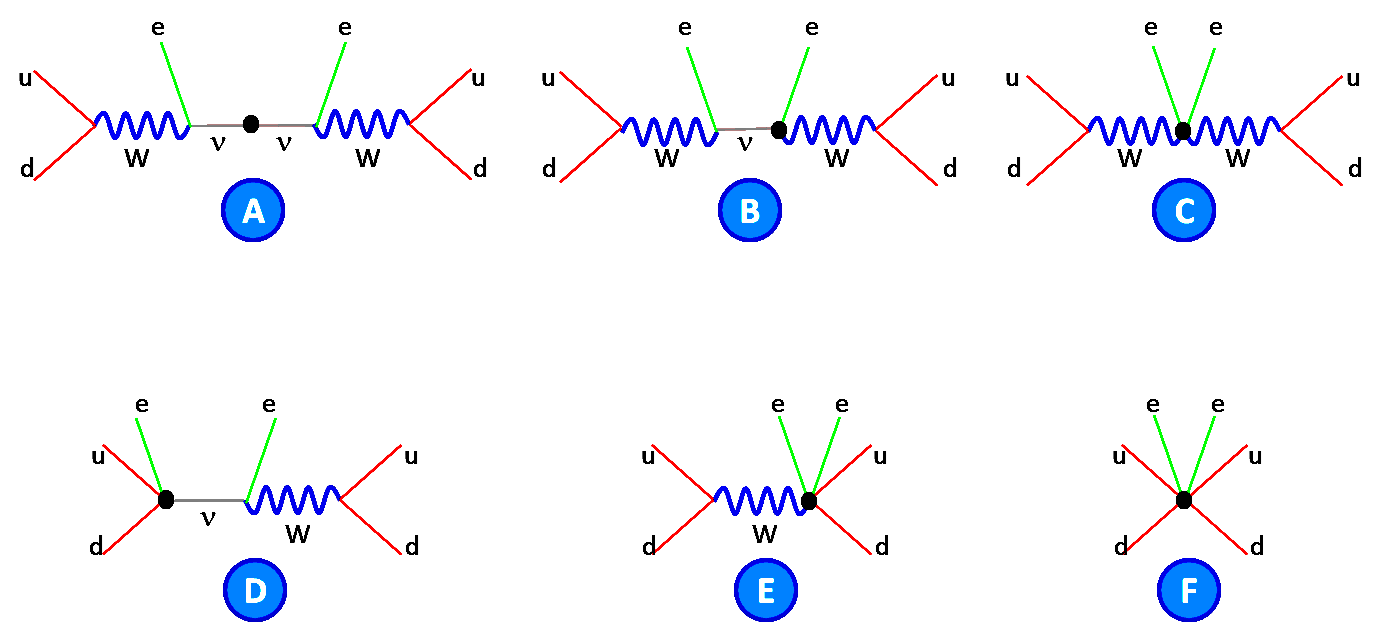}
\par\end{centering}

\caption{Classification of possible contributions to $0\nu\beta\beta$. \label{fig:0nbb-diagrams}}
\end{figure}

\begin{itemize}
\item The SM is a low-energy approximation of a more complete theory.
\item The only light particles ($m\lesssim250\,\mathrm{GeV}$) are those
of the SM.
\end{itemize}
Then, the effective Lagrangian can be written as
\begin{equation}
\mathcal{L}=\mathcal{L}_{\mathrm{SM}}+\sum_{n=5}^{\infty}\sum_{i}\left(\frac{C_{i}^{(n)}}{\Lambda^{n-4}}\mathcal{O}_{i}^{(n)}+\mathrm{h.c.}\right)\,,\label{eq:effective-langrangian}
\end{equation}
where $\mathcal{L}_{\mathrm{SM}}$ is the SM Lagrangian ($\phi$ and
$\ell$ stand for the SM Higgs and left-handed (L) lepton doublets,
$e_{R}$ the right-handed (R) charged lepton singlets, whereas flavour
indices are suppressed and $Y_{e}$ are the corresponding Yukawa couplings)
\begin{equation}
\mathcal{L}_{\mathrm{SM}}=i\overline{\ell}\slashed{D}\ell+i\overline{e_{R}}\slashed{D}e_{R}-(\overline{\ell}Y_{e}e_{R}\,\phi+\mathrm{h.c.})+\cdots\label{eq:SMLagrangian}
\end{equation}
and $\mathcal{O}_{i}^{(n)}$ are dimension-$n$ gauge invariant operators
built with SM fields, being their effects suppressed by $1/\Lambda^{n}$,
with $\Lambda$ the NP scale.

There are in general many operators $\mathcal{O}_{i}^{(n)}$ (see
for instance \cite{Babu:2001ex,Choi:2002bb,Engel:2003yr,deGouvea:2007xp}),
but if we restrict ourselves to those not involving quarks, the list
of relevant, lowest order operators, reduces to only three which,
amazingly, can be classified according to the chirality of the emitted
electrons, LL, LR, RR, and correspond to the diagrams of type $A$--$C$
in Figure~\ref{fig:0nbb-diagrams}, respectively (for details see
\cite{delAguila:2012nu}). They read

\begin{eqnarray}
\mathrm{LL\,:}\qquad\mathcal{O}^{(5)} & = & \left(\overline{\tilde{\ell}_{\alpha}}\phi\right)\left(\tilde{\phi}^{\dagger}\ell_{\beta}\right)=-v^{2}\,\overline{\nu_{\alpha\mathrm{L}}^{\mathrm{c}}}\nu_{\beta\mathrm{L}}+\dots\,,\qquad\label{eq:O5}\\
\mathrm{LR\,:\qquad}\mathcal{O}^{(7)} & = & \left(\phi^{\dagger}D_{\mu}\tilde{\phi}\right)\left(\phi^{\dagger}\overline{e_{\alpha\mathrm{R}}}\gamma^{\mu}\tilde{\ell}_{\beta}\right)=i\,\frac{gv^{3}}{\sqrt{2}}W_{\mu}^{-}\overline{e_{\alpha\mathrm{R}}}\gamma^{\mu}\nu_{\beta\mathrm{L}}^{\mathrm{c}}+\dots\,,\label{eq:O7}\\
\mathrm{RR\,:\qquad}\mathcal{O}^{(9)} & = & \overline{e_{\alpha\mathrm{R}}}e_{\beta\mathrm{R}}^{\mathrm{c}}\left(\phi^{\dagger}D\,\tilde{\phi}\right)^{2}=-\frac{g^{2}v^{4}}{2}W_{\mu}^{-}W^{-\mu}\overline{e_{\alpha\mathrm{R}}}e_{\beta\mathrm{R}}^{\mathrm{c}}+\dots\,,\label{eq:O9}
\end{eqnarray}
where $v=\langle\phi\rangle\sim174\,\mathrm{GeV}$ is the SM vacuum
expectation value (VEV) and $\tilde{\phi}=i\tau_{2}\phi^{*}$, $\tilde{\ell}=i\tau_{2}\ell^{c}$
are the conjugate SM doublets with $\tau_{2}$ the weak isospin Pauli
matrix. From the interactions (\ref{eq:O5}--\ref{eq:O9}) one can
immediately estimate the amplitudes contributing to $0\nu\beta\beta$
\begin{eqnarray}
\mathrm{LL\,:}\qquad\mathcal{A}_{0\nu\beta\beta}^{(5)} & \text{\ensuremath{\sim}} & {\displaystyle \frac{C_{ee}^{(5)}}{\text{\ensuremath{\Lambda}}p_{\mathrm{eff}}^{2}v^{2}}}\,,\label{eq:Amp5}\\
\mathrm{LR\,:}\qquad\mathcal{A}_{0\nu\beta\beta}^{(7)} & \sim & {\displaystyle \frac{C_{ee}^{(7)}}{\text{\ensuremath{\Lambda}}^{3}p_{\mathrm{eff}}v}\,,}\label{eq:Amp7}\\
\mathrm{RR\,:}\qquad\mathcal{A}_{0\nu\beta\beta}^{(9)} & \sim & {\displaystyle \frac{C_{ee}^{(9)}}{\text{\ensuremath{\Lambda}}^{5}}}\,,\label{eq:Amp9}
\end{eqnarray}
with $p_{\mathrm{eff}}\sim100\,\mathrm{MeV}$ the effective momentum
exchanged, which is estimated from complete nuclear matrix elements
calculations. $0\nu\beta\beta$ experiments%
\footnote{EXO has improved it recently in about a factor of $2$.%
} (HM,IGEX) give $T_{1/2}>1.9\times10^{25}\,\mathrm{years}$, implying 

\begin{equation}
\frac{p_{\mathrm{eff}}}{G_{F}^{2}}\,\left|\mathcal{A}_{0\nu\beta\beta}\right|\lesssim5\times10^{-9}\,.\label{eq:Amp0nbbBound}
\end{equation}
Imposing this bound on the different terms, we obtain

\begin{eqnarray}
\mbox{\ensuremath{\mathrm{LL\,:}}} & \quad\phantom{\frac{\Lambda}{\Lambda}} & {\Lambda>10^{11}}\,|C_{ee}^{(5)}|\;\mathrm{{TeV}\,,}\label{eq:Lambda5}\\
\mathrm{LR\,:} & \quad\phantom{\frac{\Lambda}{\Lambda}} & {\Lambda>106}\,|C_{ee}^{(7)}|^{\nicefrac{1}{3}}\;\mathrm{{TeV}\,,}\label{eq:Lambda7}\\
\mathrm{RR\,:} & \quad\phantom{\frac{\Lambda}{\Lambda}} & {\Lambda>2.7}\,|C_{ee}^{(9)}|^{\nicefrac{1}{5}}\;\mathrm{{TeV}\,.}\label{eq:Lambda9}
\end{eqnarray}
By using detailed nuclear matrix elements these estimates do not substantially
change.

\subsection{Contribution to $\nu$ masses\label{sub:Contribution-to-numasses}}

Once LN is violated by the operators ${\displaystyle \mathcal{O}^{(n)}}$,
neutrino masses (that is $\mathcal{O}^{(5)}$) will be sooner or later
generated. Thus, $\mathcal{O}^{(7)}$ will give one-loop contributions
to neutrino masses while the first contributions to neutrino masses
from $\mathcal{O}^{(9)}$ will arise at two loops. Although these
contributions cannot be calculated precisely in the effective theory,
they can be estimated by naive dimensional analysis. In Table~\ref{tab:Contribution-to-neutrino}
we present the relevant diagrams and estimates for the masses. It
is important to notice the presence of charged fermion mass factors,
$m_{a}$. As emphasized in the introduction, in the $m_{a}\rightarrow0$
limit, lepton numbers carried by $\nu_{L}$ and $e_{L}$ cannot be
linked to that carried by $e_{R}$. Hence, one mass insertion is needed
to generate neutrino masses from $\mathcal{O}^{(7)}$ (which involves
$e_{L}$ and $e_{R}$) and two mass insertions to generate them from
$\mathcal{O}^{(9)}$ (which involves two $e_{R}$). The hierarchy
of charged lepton masses then translates into a quite characteristic
structure of neutrino masses with interesting phenomenological consequences.
We will discuss them in the framework of specific models where all
the coefficients can be precisely calculated. 

\begin{table}
\begin{tabular}{>{\centering}p{20mm}>{\centering}m{60mm}>{\centering}p{55mm}}
LL~: & \includegraphics[width=0.18\columnwidth]{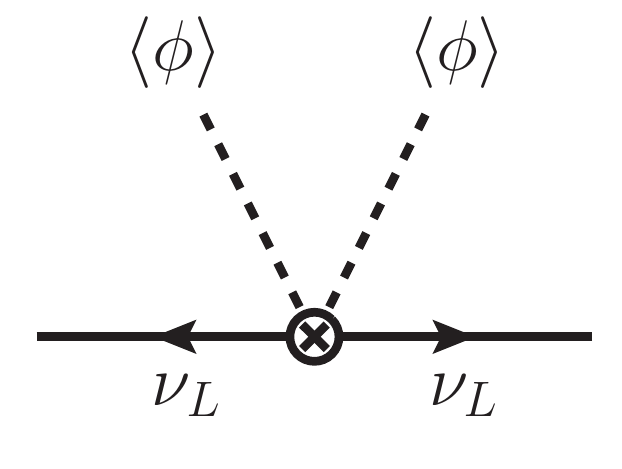} & ${\displaystyle (m_{\nu})_{ab}\sim\frac{v^{2}}{\Lambda}C_{ab}^{(5)}}$\tabularnewline
LR~: & \includegraphics[width=0.33\columnwidth]{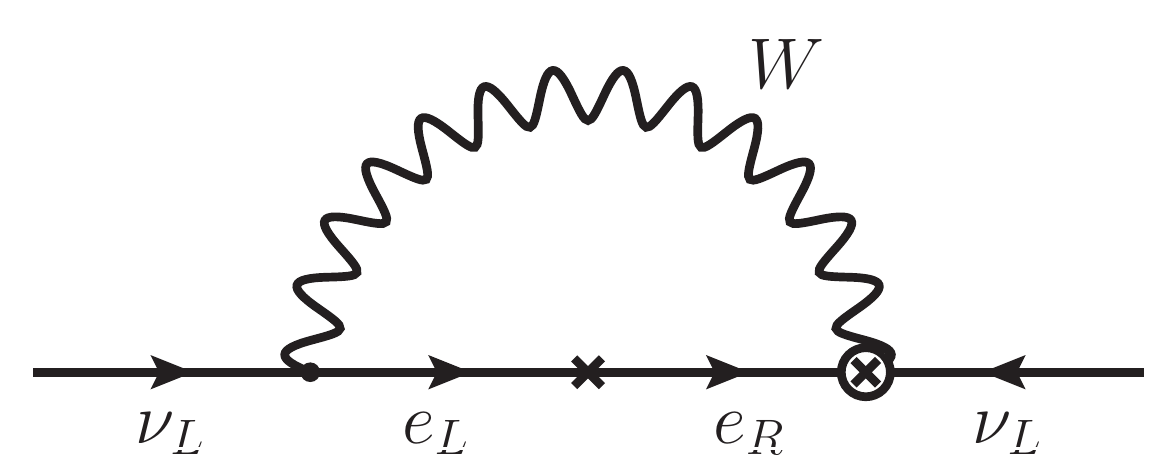}

~ & ${\displaystyle (m_{\nu})_{ab}\sim\frac{v}{16\pi^{2}\Lambda}\left(m_{a}C_{ab}^{(7)}+m_{b}C_{ba}^{(7)}\right)}$\tabularnewline
RR~: & \includegraphics[width=0.4\columnwidth]{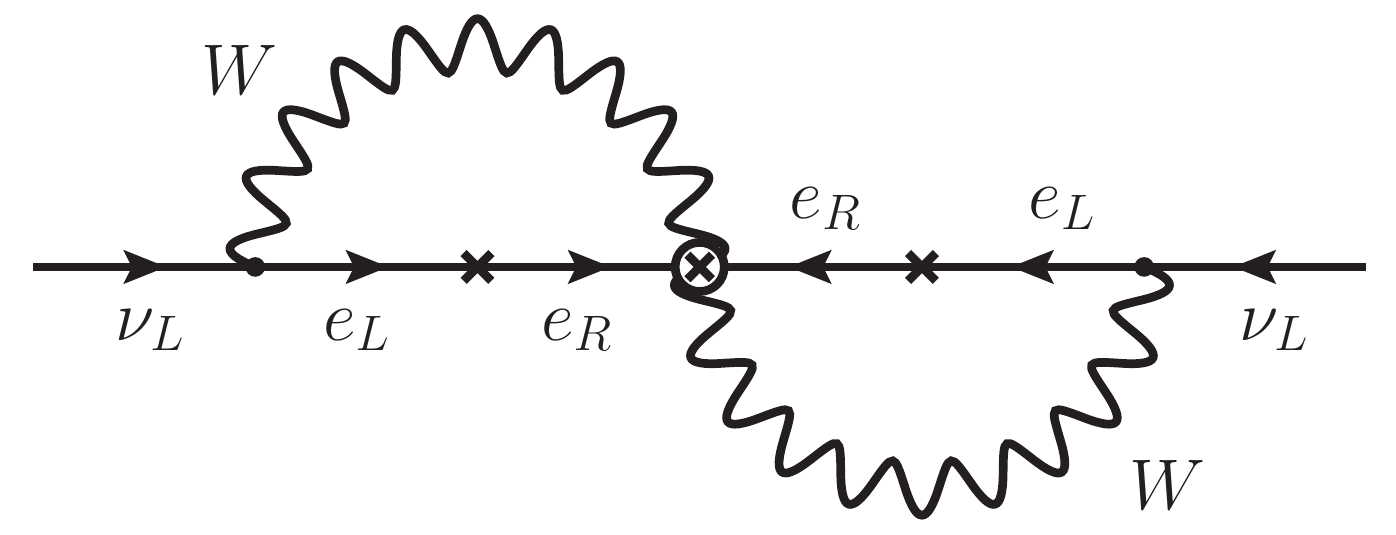} & ${\displaystyle (m_{\nu})_{ab}\sim\frac{1}{\text{(16\ensuremath{\pi}}^{2})^{2}\Lambda}m_{a}C_{ab}^{(9)}m_{b}}$\tabularnewline
\end{tabular}

\caption{Contribution to neutrino masses from the relevant operators giving
rise to $0\nu\beta\beta$.\label{tab:Contribution-to-neutrino}}

\end{table}

Assuming that the dominant contribution to neutrino masses comes from
the operators ${\displaystyle \mathcal{O}^{(5-9)}}$ and the diagrams
in Table \ref{tab:Contribution-to-neutrino}, we can correlate them
and the $0\nu\beta\beta$ amplitudes

\begin{eqnarray}
\mathrm{LL}\,:\qquad\mathcal{A}_{0\nu\beta\beta}^{(5)} & \text{\ensuremath{\propto}} & (m_{\nu})_{ee}\,,\label{eq:Amp2mnu5}\\
\mathrm{LR}\,:\qquad\mathcal{A}_{0\nu\beta\beta}^{(7)} & \propto & (m_{\nu})_{ee}{(4\pi)^{2}\frac{v^{2}}{\Lambda^{2}}\frac{p_{\mathrm{eff}}}{m_{e}}}\,,\label{eq:Amp2mnu7}\\
\mathrm{RR\,:}\qquad\mathcal{A}_{0\nu\beta\beta}^{(9)} & \propto & (m_{\nu})_{ee}\left({(4\pi)^{2}\frac{v^{2}}{\Lambda^{2}}\frac{p_{\mathrm{eff}}}{m_{e}}}\right)^{2}\,.\label{eq:Amp2mnu9}
\end{eqnarray}
Thus, in general, the standard Majorana neutrino mass contribution
to $0\nu\beta\beta$ in Figures \ref{fig:numassTo0nbb} and \ref{fig:0nbb-diagrams}A,
will be always present together with the new contributions given by
the effective operators in Figure \ref{fig:0nbb-diagrams}B-F. The
interesting situation occurs when the new contributions dominate over
the standard one exchanging a massive Majorana neutrino, otherwise
the standard analysis applies. Thus, LR and RR contributions are larger
when

\begin{equation}
\Lambda<4\pi v\sqrt{\frac{p\mathrm{_{eff}}}{m_{e}}}\sim30\,\mathrm{TeV}\,\label{eq:LambdaBound}
\end{equation}
and, therefore, the NP scale is relatively low, and perhaps accessible
to LHC.

\section{Renormalizable Completions\label{sec:Renormalizable-Completions}}

The effective Lagrangian approach is very general and, as seen, just
using it one can grasp the main consequences of the different scenarios.
However, sometimes it is quite difficult to realize a specific effective
Lagrangian with a renormalizable model. Therefore, it is important
to check that these effective theories can arise from a consistent
renormalizable model and that the estimates obtained in the effective
Lagrangian approach are correct. Moreover, in our case, since the
scale of NP is not extremely large, perhaps the new particles are
accessible at LHC and/or provide interesting effects in lepton flavour
violation (LFV) processes, providing additional tests of the proposed
mechanisms for $0\nu\beta\beta$ . Of course, one needs explicit models
to address this.\textsc{}
\begin{figure}
\begin{centering}
\includegraphics[width=0.8\columnwidth]{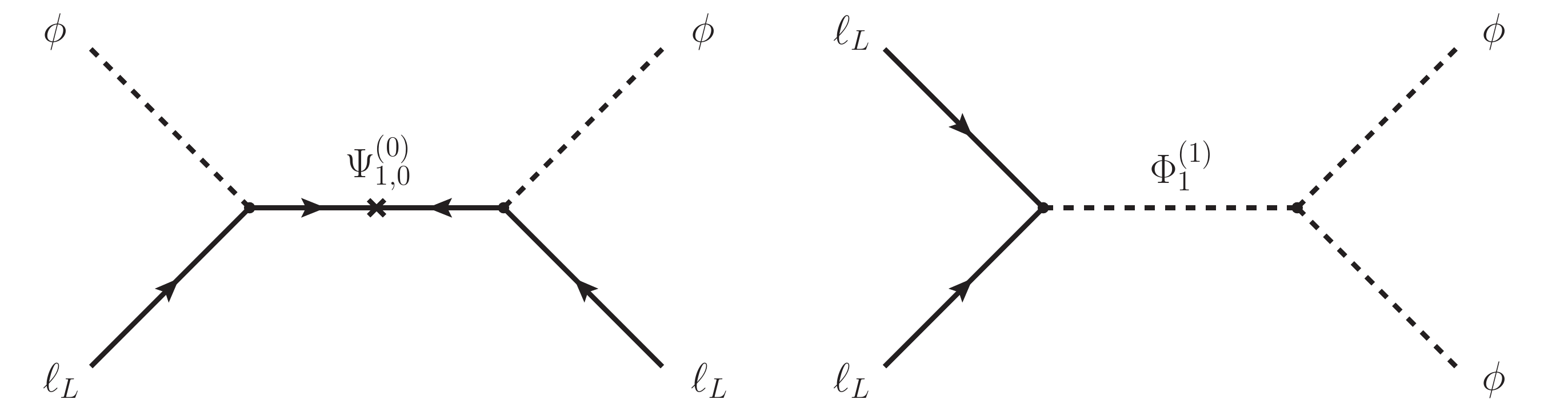}
\par\end{centering}

\caption{The three see-saw mechanisms generating the operator $\mathcal{O}^{(5)}$
(LL). \label{fig:The-three-see-saw}}
\end{figure}

The operators $\mathcal{O}{}^{(5)}$, $\mathcal{O}^{(7)}$, $\mathcal{O}^{(9)}$
can be obtained from renormalizable interactions by adding a variety
of new particles. We will denote by $\Phi_{I}^{(Y)}$ scalars with
hypercharge $Y$ and weak isospin $I$. Similarly $\Psi_{I}^{(Y)}$
will be spin $1/2$ particles and $X_{I}^{(Y)}$ vector bosons. The
different ways of obtaining the operator $\mathcal{O}^{(5)}$ (LL)
at tree level have been extensively studied and constitute the three
see-saw mechanisms which are represented diagrammatically in Figure~\ref{fig:The-three-see-saw}:
(left) exchange of a neutral fermion $\Psi_{0}^{(0)}$ (type I see-saw)
or of a fermion triplet $\Psi_{1}^{(0)}$ (type III see-saw); (right)
exchange of a scalar triplet $\Phi_{1}^{(1)}$ (type II see-saw)\textsc{.}
For $\mathcal{O}^{(7)}$~~(LR) and $\mathcal{O}^{(9)}$~~(RR)
there are many more possibilities (they have been classified in \cite{delAguila:2012nu}).
For instance, $\mathcal{O}^{(7)}$ can be obtained adding $\{\Psi_{1/2}^{(1/2)},\,\Phi_{1}^{(1)}\}\,;\,\{\Psi_{0}^{(0)},\,\Phi_{0}^{(1)}\}\,,\cdots$,
and $\mathcal{O}^{(9)}$ results from integrating out $\{\Phi_{0}^{(2)},\,\Phi_{1}^{(1)}\}\,;\,\{\Psi_{1}^{(0)},\,\Phi_{1}^{(1)}\}\,,\cdots$.
In the case that $0\nu\beta\beta$ is induced by $\mathcal{O}^{(7,9)}$
we would like to forbid tree-level Majorana neutrino masses because
otherwise they will dominate $0\nu\beta\beta$. Models can be supplemented
with additional global continuous or discrete symmetries to forbid
$\mathcal{O}^{(5)}$. In the following we will present some examples
of renormalizable theories giving rise to these operators together
with their characteristic phenomenology.

\subsection{An example of LR model}

We start from the set $\{\Psi_{1/2}^{(1/2)},\,\Phi_{1}^{(1)}\}$.
In order to simplify the notation we define $\Phi_{1}^{(1)}\equiv\chi$,
a scalar isotriplet of hypercharge $1$, and $\Psi_{1/2}^{(1/2)}\equiv L^{c}=L_{L}^{c}+L_{R}^{c}$,
a lepton isodoublet of hypercharge $1/2$ (in terms of its left-handed
and right-handed components). A simple way to insure the decoupling
of the heavy physics is to assume, as we do, that the heavy fermions
are vector-like. This particle content is sufficient to generate $\mathcal{O}^{(7)}$
at tree level, and it is not hard to convince oneself that the relevant
graphs must involve the couplings $e_{R}\phi\tilde{L}$, $\ell L\chi$
and $\phi^{\dagger}\phi^{\dagger}\chi$. However, such a model also
allows the coupling $\ell\ell\chi$ and will then generate $\mathcal{O}{}^{(5)}$
at tree level through the standard type-II see-saw diagram. In order
to avoid this we impose a discrete $Z_{2}$ symmetry under which $\chi$
and $L$ are odd and $\ell$ is even; unfortunately, this symmetry
also forbids the $\phi^{\dagger}\phi^{\dagger}\chi$ vertex. In order
to overcome this difficulty we assume the presence of two light scalar
doublets $\phi,~\phi'$, which are even and odd under $Z_{2}$, respectively.
Moreover, in order to accommodate a generic neutrino mass matrix and
also allow for flavour symmetries treating the three families on the
same footing, we will assume the presence of 3 heavy vector-like fermion
doublets $L_{a},~a=1,2,3$. The complete list of new fields is given
in Table \ref{tab:modelfieldsLR}. The Lagrangian will include all
renormalizable couplings preserving these symmetries, noting that
the SM fields transform trivially under $Z_{2}$.

\begin{table}
\centering{}%
\begin{tabular}{l|cccc}
$ $  & \quad{}$L_{La}$ \quad{} & \quad{}$L_{Ra}$ \quad{} & \quad{}$\chi$ \quad{} & \quad{}$\phi^{\prime}$ \quad{}\tabularnewline
\hline 
$SU(2)_{L}$  & $\frac{1}{2}$  & $\frac{1}{2}$  & 1  & $\frac{1}{2}$ \tabularnewline
$U(1)_{Y}$  & $-\frac{1}{2}$  & $-\frac{1}{2}$  & 1  & $\frac{1}{2}$ \tabularnewline
$Z_{2}$  & $-$  & $-$  & $-$  & $-$ \tabularnewline
\end{tabular}\caption{Quantum number assignment for the extra fields in a model realizing
the $\mathcal{O}^{(7)}$ operator.\textsc{\label{tab:modelfieldsLR}}}
\end{table}

Thus, the heavy lepton Lagrangian reads 
\begin{equation}
\mathcal{L}_{\mathrm{H}}^{L}=\overline{L_{a}}(i\slashed D-M_{a})L_{a}+\{y_{ab}^{e}\overline{L_{aL}}\phi^{\prime}e_{bR}+y_{ab}^{\nu}\overline{\tilde{L}_{aL}}\chi\ell_{b}+\mathrm{h.c.}\}\ ,\label{eq:heavyLR}
\end{equation}
where we assumed, without loss of generality, that the heavy lepton
mass matrix is diagonal. Once $\phi'$ and $\chi$ acquire VEVs the
light $e_{R}$, $\ell$ leptons mix with the $L_{a}$. The low-energy
effects of such mixings will be proportional to $y_{ab}^{e}\vevof{\phi'}/M_{a}$
or $y_{ab}^{\nu}\vevof{\chi}/M_{a}$ and can be made as small as experimentally
required by increasing the heavy masses $M_{a}$, reducing the couplings
$y_{ab}^{e,\nu}$, or the VEVs $\vevof{\phi^{\prime}},\vevof{\chi}$
(for a recent review on vector-like leptons see, for instance, \cite{delAguila:2008pw}).
LFV effects can be further suppressed by assuming that the light charged
leptons, which get their masses through the SM Higgs mechanism, are
aligned along the heavy flavours. This corresponds to taking $y_{ab}^{e}$
diagonal, which may be natural in a larger model.

The scalar potential can be easily arranged to insure a minimum where
$\vevof{\phi}\gg\ \vevof{\phi^{\prime}},\vevof{\chi}\ \neq\ 0$, with
$\vevof{\chi}\simeq-\mu^{*}\vevof{\phi^{\prime}}\vevof{\phi}/m_{\chi}^{2}$,
$\mu$ the trilinear $\phi^{\dagger}\chi\tilde{\phi'}$ coupling and
$m_{\chi}$ the isotriplet mass (in order to satisfy the limit from
electroweak precision data \cite{Nakamura:2010zzi,delAguila:2008ks}
we require $\vevof{\chi}\lesssim2$ GeV). We assume negative mass
terms for $\phi$ and $\phi^{\prime}$ to trigger the corresponding
VEVs, whereas $\chi$ gets a VEV through its mixing with the scalar
isodoublets. Otherwise, dimensional couplings in the potential are
typically of electroweak order, except for new scalar masses that
may be larger. Dimensionless ones stay perturbative, in general ranging
within an $\alpha_{EM}\sim10^{-2}$ factor. Note that in this model
LN is explicitly broken by (renormalizable) terms in the scalar potential,
in particular by the $(\phi^{\dagger}\phi^{\prime})^{2}$ term. 

It is important to remark that in this model there cannot be tree-level
neutrino masses because: (i) $L_{a}$ are doublets and therefore,
they cannot mediate the see-saw of types I-III. (ii) There is no coupling
$\chi\ell\ell$ due to the discrete symmetry and hence, no type II
see-saw contributions.

\subsubsection{The LR operator for $0\nu\beta\beta$}

\begin{figure}
\begin{centering}
\includegraphics[width=0.5\columnwidth]{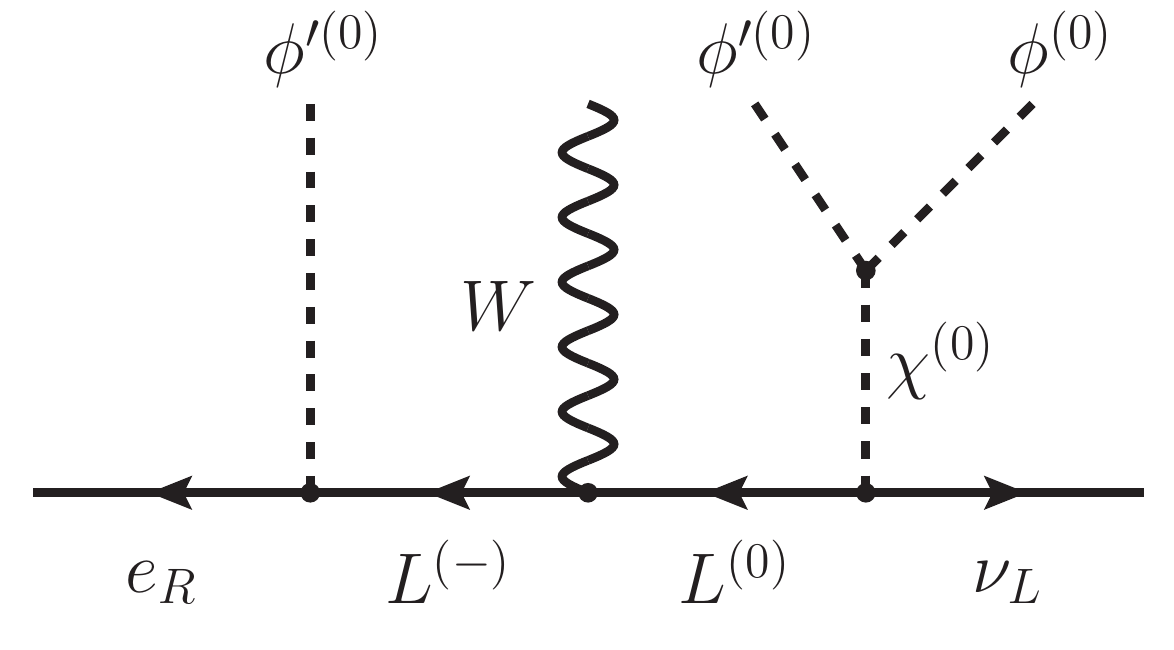} 
\par\end{centering}

\caption{Tree-level diagram contributing to $\mathcal{O}^{(7)}$ in the model
proposed.\textsc{ \label{fig:O7generation}}}
\end{figure}

Given the couplings of the model one can evaluate $C_{ab}^{(7)}$
by using the diagram in Figure \ref{fig:O7generation},

\begin{equation}
\frac{C_{ab}^{(7)}}{\Lambda^{3}}=-i\frac{\mu y_{ca}^{e*}y_{cb}^{\nu*}}{m_{\chi}^{2}M_{c}^{2}}\,,\label{eq:C7}
\end{equation}
where all masses in the $L_{c}$ and $\chi$ multiplets are taken
equal.

\subsubsection{The neutrino mass}

\begin{figure}
\begin{centering}
\includegraphics[width=0.5\columnwidth]{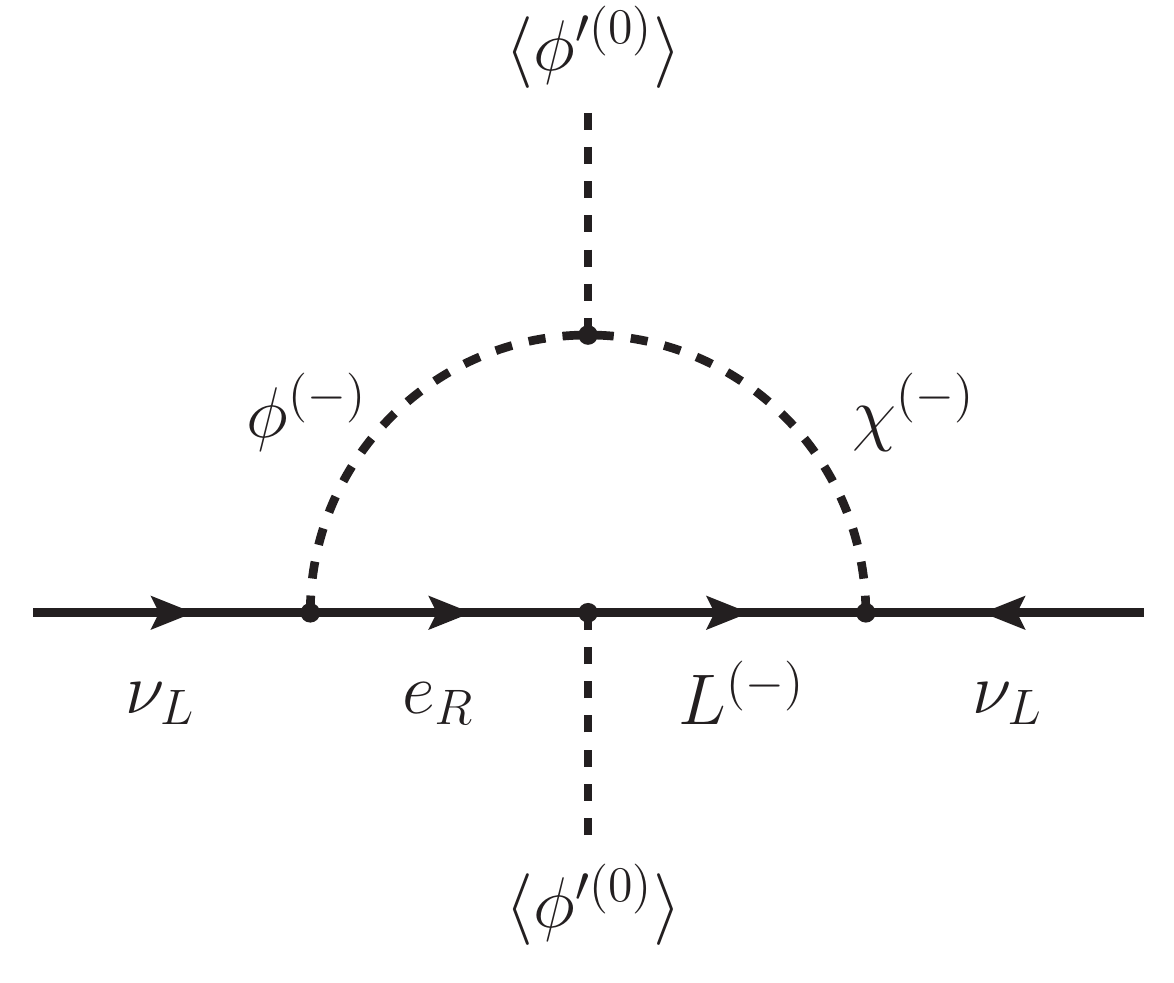} 
\par\end{centering}

\caption{Leading one-loop contribution in the Feynman gauge to neutrino masses
in the model generating $\mathcal{O}^{(7)}$. \label{fig:neutrinomassesLRmodel}}
\end{figure}

Although neutrinos are massless at tree level, as expected from the
discussion in Section \ref{sub:Contribution-to-numasses}, they will
acquire a mass at one loop. The dominant contribution can be obtained
from the diagram in Figure~\ref{fig:neutrinomassesLRmodel}. This
is quite different from that used to estimate the mass in the effective
theory. This is due to the intricacies of gauge symmetry, but the
dominant piece in the limit of small gauge couplings, $g\rightarrow0$,
can be obtained by using only Yukawa couplings as in Figure~\ref{fig:neutrinomassesLRmodel}.
The result being 

\begin{equation}
(m_{\nu})_{ab}\simeq\frac{v^{\prime\,2}\mu}{32\pi^{2}v}\left(m_{a}y_{ca}^{e*}y_{cb}^{\nu*}+m_{b}y_{cb}^{e*}y_{ca}^{\nu*}\right)\frac{1}{M_{c}^{2}-m_{\chi}^{2}}\log\frac{M_{c}^{2}}{m_{\chi}^{2}}\,,\label{eq:mnuLRmodel}
\end{equation}
where $v^{\prime}=\vevof{\phi^{\prime}}$ and we have assumed that
all other masses are much smaller than $M_{c}$ and $m_{\chi}$. Thus,
with only one heavy lepton doublet the neutrino mass matrix has at
most rank 2. With two heavy lepton doublets all three light neutrinos
can be massive. \textsc{}

Comparing eqs.~\eqref{eq:mnuLRmodel} and eq.~\eqref{eq:C7} with
the neutrino mass formula in Table~\ref{tab:Contribution-to-neutrino}
for the LR case we see that the general formula roughly applies (take
for instance $v'=v$ and $\Lambda=M_{c}=m_{\chi}$), but by taking
different values for the VEVs and the masses of the new particles
one can change substantially the numerical values. Since the neutrino
masses are generated at one loop and are suppressed by only one charged
lepton mass they will tend to be too large for masses $m_{\chi},M_{c}$
just above the electroweak scale and Yukawa couplings of order one.
However, by playing with the ratios $v^{\prime}/v$ and $m_{\chi}/M_{c}$
one can obtain additional suppression factors for neutrino masses
relative to $0\nu\beta\beta$.

\subsubsection{Relevant Phenomenology}

The phenomenology of the model is very rich and can be summarized
as follows:

By adjusting $v^{\prime}/v$, $m_{\chi}/M_{c}$ and the couplings
one can enhance the tree-level contributions to $0\nu\beta\beta$
enough to be at the reach of the present round of experiments. On
the other hand, the neutrino mass matrix has a very characteristic
structure, eq.~\eqref{eq:mnuLRmodel}, but still can accommodate
the observed neutrino spectrum. The corresponding neutrino mass constraints
together the requirement of a large $0\nu\beta\beta$ favour small
couplings and relatively light new particles. 

Thus, $L_{a}$, $\chi$ and $\phi^{\prime}$ can be discovered at
LHC if light enough ( $\lesssim800\,\mathrm{GeV}$), but it depends
on the details of the couplings and decay channels. The scalar triplet
contains a doubly-charged scalar which will be easy to see if it decays
mainly to $e,\mu$, but it will be more complicated to discover if
it decays to $\tau$, $W$ or singly-charged scalars. The production
of vector-like leptons $L_{a}$ at LHC has been previously studied
\cite{delAguila:1989rq}, with the general conclusion that they can
be detected provided their masses are below $\sim850$~GeV for a
center of mass (CM) energy of $14$~TeV and an integrated luminosity
of $100$~fb$^{-1}$ \cite{AguilarSaavedra:2009ik}. The LHC reach
reduces to $\sim350$~GeV for heavy leptons mainly decaying into
taus \cite{delAguila:2010es}.

LFV can be always made small at tree level even though once $\phi'$
and $\chi$ acquire VEVs light $e_{R}$, $\ell$ leptons mix with
the heavy ones $L_{a}$. Such mixings and the corresponding phenomenology
of heavy vector-like lepton doublets were analyzed long ago in different
contexts (for a review see \cite{Raidal:2008jk,Feldmann:2011zh};
for updated limits see \cite{delAguila:2008pw}). The low-energy effects
of these mixings are proportional to $y_{ab}^{e}\vevof{\phi'}/M_{a}$
or $y_{ab}^{\nu}\vevof{\chi}/M_{a}$, and can be made as small as
experimentally required by increasing the heavy masses $M_{a}$, reducing
the couplings $y_{ab}^{e,\nu}$, or the VEVs $\vevof{\phi^{\prime}},\vevof{\chi}$.
LFV effects can be further suppressed by assuming that the light charged
leptons, which get their masses through the SM Higgs mechanism, are
aligned along the heavy flavors. This corresponds to taking $y_{ab}^{e}$
diagonal, which may be natural in a larger model.

At one loop LFV through the exchange of heavy leptons and bosons provides
the most stringent constraints on this model, but they can be avoided.
The most restrictive processes are those involving the muon to electron
transition. Then, one can, to a large extent, apply the conclusions
from related analyses for the Littlest Higgs model with T-parity \cite{Blanke:2007db,delAguila:2008zu,Goto:2010sn,delAguila:2010nv};
the general conclusion is that the heavy flavors must be aligned with
the light charged leptons with a precision better than $1-10$ \%
for heavy masses of ${\cal O}({\rm TeV})$.

More details on the phenomenology of this model can be found in \cite{delAguila:2012nu}.

\subsection{An example of RR model}

In order to generate $\mathcal{O}^{(9)}$ we can add the scalars $\{\Phi_{0}^{(2)},\,\Phi_{1}^{(1)}\}$
to the SM. For easy notation we will name the doubly-charged scalar
$\text{\ensuremath{\Phi}}_{0}^{(2)}=\kappa$ and the triplet scalar
$\Phi_{1}^{(1)}=\chi$. Moreover, we will impose a discrete $Z_{2}$
symmetry in order to forbid the $\chi\ell\ell$ coupling, which would
provide tree-level 
\begin{table}[h]
\begin{centering}
\begin{tabular}{l|ccc}
$ $  & \quad{}$\chi$ \quad{} & \quad{}$\kappa$ \quad{} & \quad{}$\sigma$ \quad{}\tabularnewline
\hline 
$SU(2)_{L}$  & $1$  & $0$  & $0$ \tabularnewline
$U(1)_{Y}$  & $1$  & $2$  & $0$ \tabularnewline
$Z_{2}$  & $-$  & $+$  & $-$ \tabularnewline
\end{tabular}
\par\end{centering}

\caption{New fields and their quantum numbers in a model realizing the $\mathcal{O}^{(9)}$
operator.\textsc{\label{tab:modelfieldsRR}}}
\end{table}
neutrino masses once the triplet develops a VEV. This symmetry can
be implemented adding a real scalar $\sigma$, odd under $Z_{2}$.
This scalar is not really necessary if one allows for a soft breaking
of this discrete symmetry, what provides a simpler variation of the
model. Other variations are discussed in \cite{delAguila:2011gr}
but most of their phenomenology is shared by the model discussed here.
The spectrum of new particles and their quantum numbers are displayed
in Table~\ref{tab:modelfieldsRR}, allowing from the new terms in
the Lagrangian
\begin{equation}
\mathcal{L}=g_{\alpha\beta}\,\overline{e_{\alpha\mathrm{R}}}^{\mathrm{c}}e_{\beta\mathrm{R}}\,\kappa-\mu_{\kappa}\,\kappa\mathrm{Tr}\left\{ \chi^{\dagger}\chi^{\dagger}\right\} -\lambda_{6}\,\sigma\,\phi^{\dagger}\chi\tilde{\phi}+\cdots\,.\label{eq:LagrangianRR}
\end{equation}
This model does not have tree-level neutrino masses because: (i) There
are no new fermions in the spectrum and therefore, there cannot be
type I-III see-saw neutrino mass contributions. (ii) There is no coupling
$\chi\ell\ell$ due to the discrete symmetry and hence, type II see-saw
neutrino masses do not arise either.

\subsubsection{The RR operator for $0\nu\beta\beta$}

From the Lagrangian in eq.~\eqref{eq:LagrangianRR} and the diagram
in Figure~\ref{fig:O9generation}, one can obtain the $\mathcal{O}^{(9)}$
coefficient in the corresponding effective Lagrangian

\begin{center}
\begin{figure}
\begin{centering}
\includegraphics[width=0.5\textwidth]{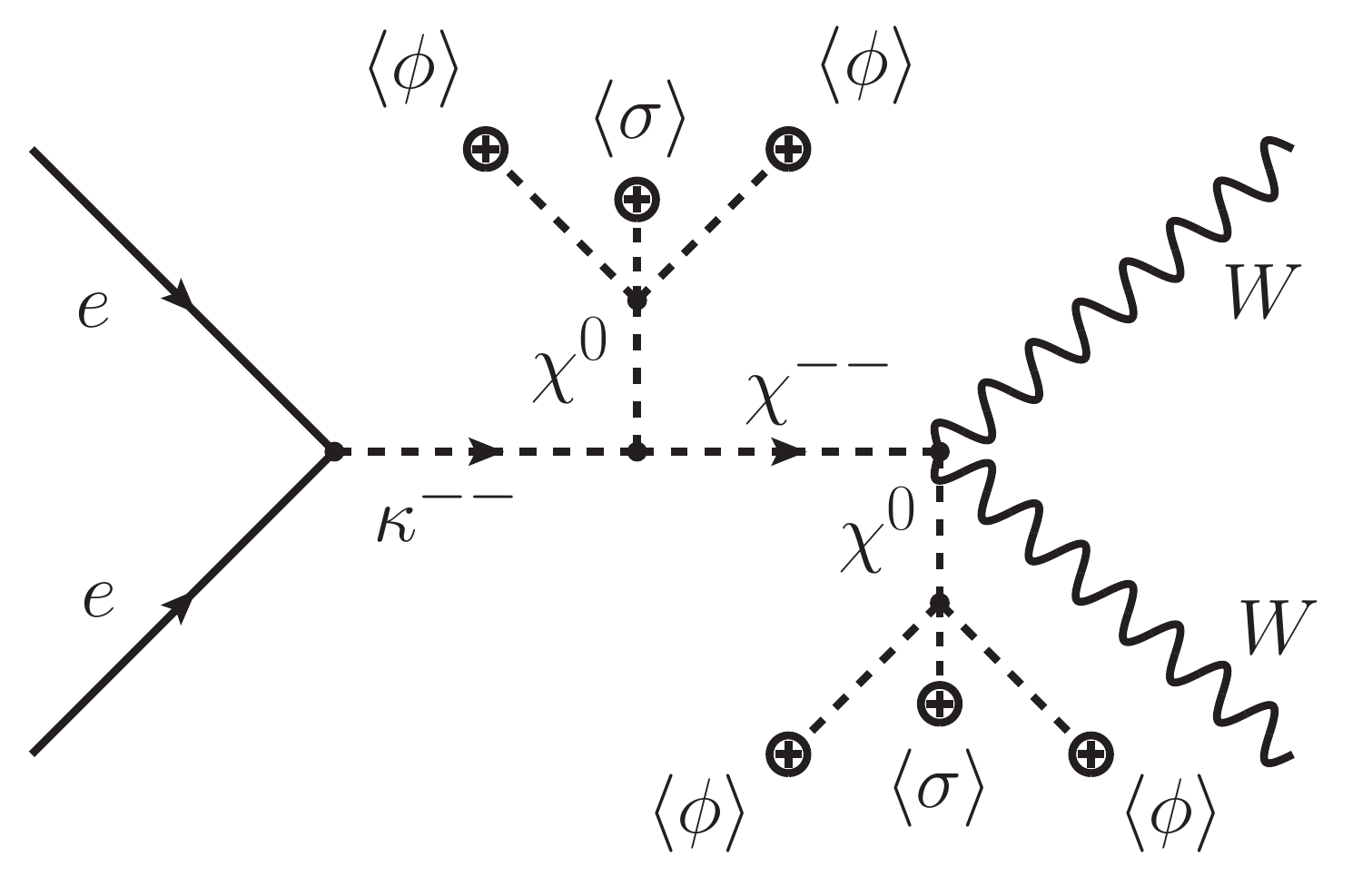}
\par\end{centering}

\caption{Tree-level diagram contributing to $\mathcal{O}^{(9)}$ in the model
proposed. \label{fig:O9generation}}
\end{figure}

\par\end{center}

\begin{equation}
\frac{C_{ab}^{(9)}}{\Lambda^{5}}=-i\frac{4v_{\chi}^{2}\mu_{\kappa}}{m_{\kappa}^{2}m_{\chi}^{2}v^{4}}g_{ab}^{*}\,,\label{eq:C9}
\end{equation}
where the result is expressed in terms of the SM VEV, $v$, and the
triplet one, $v_{\chi}\approx-\lambda_{6}\langle\sigma\rangle\langle\phi\rangle^{2}/m_{\chi}^{2}$,
as well as the new scalar masses.

\subsubsection{The neutrino mass}

As expected, in this case the neutrino mass arises at two loops. The
dominant diagram is 
\begin{figure}
\begin{centering}
\includegraphics[width=0.5\textwidth]{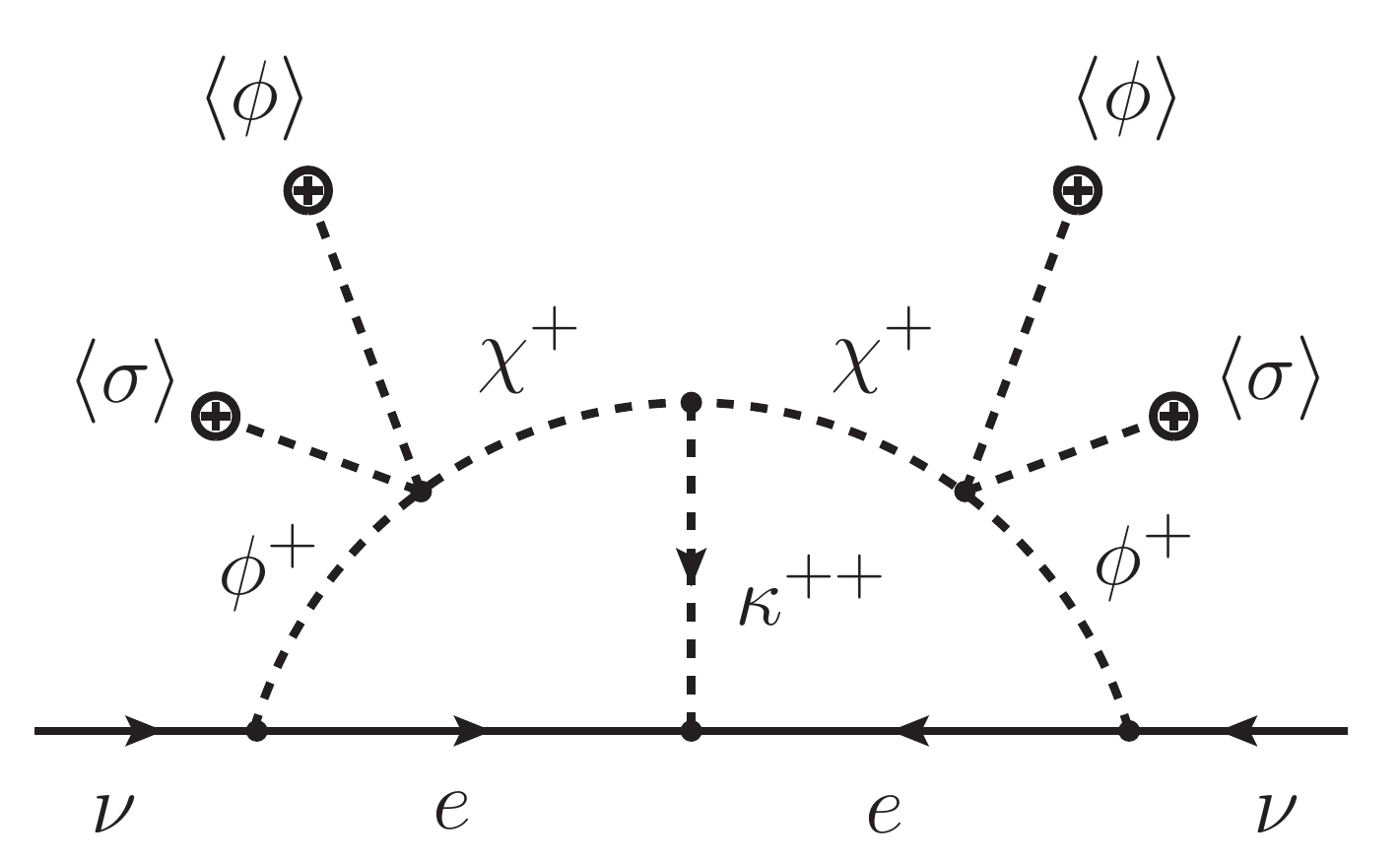}
\par\end{centering}

\caption{Leading two-loop contribution in the Feynman gauge to neutrino masses
in the model generating $\mathcal{O}^{(9)}$. \label{fig:neutrinomassesRRmodel}}
\end{figure}
depicted in Figure~\ref{fig:neutrinomassesRRmodel}. The complete
calculation, including gauge contributions, can be found in \cite{delAguila:2011gr}:

\begin{equation}
\left(m_{\nu}\right)_{\alpha\beta}=\frac{\mu_{\kappa}v_{\chi}^{2}}{2(2\pi)^{4}v^{4}}\, m_{\alpha}g_{\alpha\beta}^{*}m_{\beta}\, I_{\nu}\,,\label{eq:neutrinomassRR}
\end{equation}
where $I_{\nu}$ is an order 1 dimensionless function of the new scalar
masses and the $W$ mass. The form of $\left(m_{\nu}\right)_{\alpha\beta}$
agrees with the estimate in Table~\ref{tab:Contribution-to-neutrino}
but contains more parameters which can be used to adjust the neutrino
mass matrix elements to fit the observed neutrino spectrum, while
keeping $0\nu\beta\beta$ at a measurable level.

\subsubsection{LFV in the RR model}

\begin{figure}
\begin{centering}
\includegraphics[width=0.5\columnwidth]{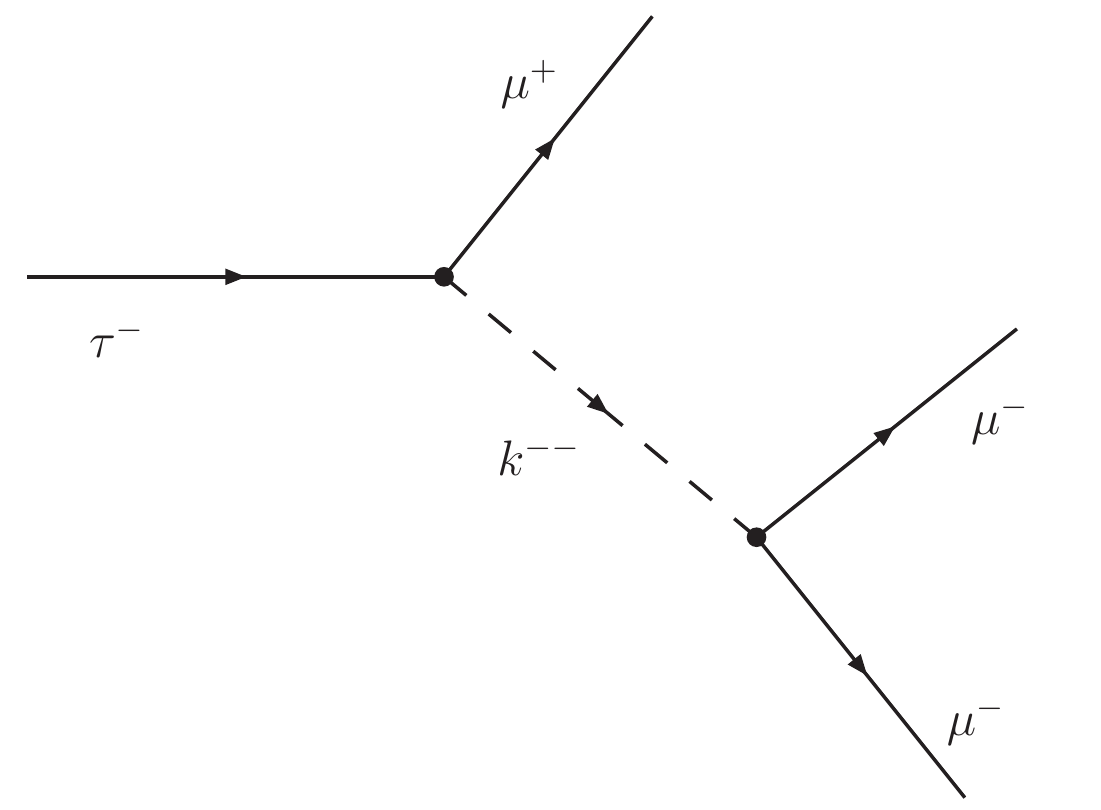}
\par\end{centering}

\caption{Diagram giving LFV decays.\label{fig:LFV}}
\end{figure}

The exchange of the doubly-charged scalar singlet gives tree-level
three-body decays for the SM charged leptons, which do not conserve
family lepton number ($\ell_{a}^{-}\rightarrow\ell_{b}^{+}\ell_{c}^{-}\ell_{d}^{-}$
in Figure~\ref{fig:LFV} and similar diagrams for $\mu^{-}$ decay).
There are very strong experimental limits on the branching ratios
of these processes, and since their amplitudes are proportional to
the product $g_{ab}g_{cd}^{*}$, one can set stringent constraints
on these couplings. 
From $\mathrm{BR}(\mu^{-}\rightarrow e^{+}e^{-}e^{-})<1.0\times10^{-12}$
one obtains 
\begin{equation}
|g_{\mu e}g_{ee}^{*}|<2.3\times10^{-5}\,(m_{\kappa}/\mathrm{TeV})^{2}\,,\label{eq:mu3e-bound}
\end{equation}
whereas $\mathrm{BR}(\tau^{-}\rightarrow e^{+}\mu^{-}\mu^{-})<1.7\times10^{-8}$
implies
\begin{equation}
|g_{\tau e}g_{\mu\mu}^{*}|<0.007\,(m_{\kappa}/\mathrm{TeV})^{2}\,.\label{eq:tau-bound}
\end{equation}
There are also bounds on other muon and tau LFV decay channels ($\mu\rightarrow e\gamma,\tau\rightarrow e\gamma,$$\tau\rightarrow eee$,
$\mu$-$e$ conversion in nuclei, $\cdots$), but for the time being
these two processes provide the strongest constraints on the model
if one fits the observed neutrino masses allowing for large additional
contributions to $0\nu\beta\beta$.

\subsubsection{Indirect constraints on doubly-charged scalars}

Since we want to adjust the neutrino masses, have a sizable $0\nu\beta\beta$
and satisfy LFV constraints, the model is quite restricted unless
one goes to parameter regions where perturbativity does not hold.
In particular the doubly-charged scalar masses are quite constrained:
if the scalar masses are too large then $0\nu\beta\beta$ and neutrino
masses will be too small and if the scalar masses are too small the
neutrino masses will be too large, LFV limits will be problematic,
as well as the bounds from LEP and LHC. 

\begin{figure}
\begin{centering}
\includegraphics[width=0.5\textwidth]{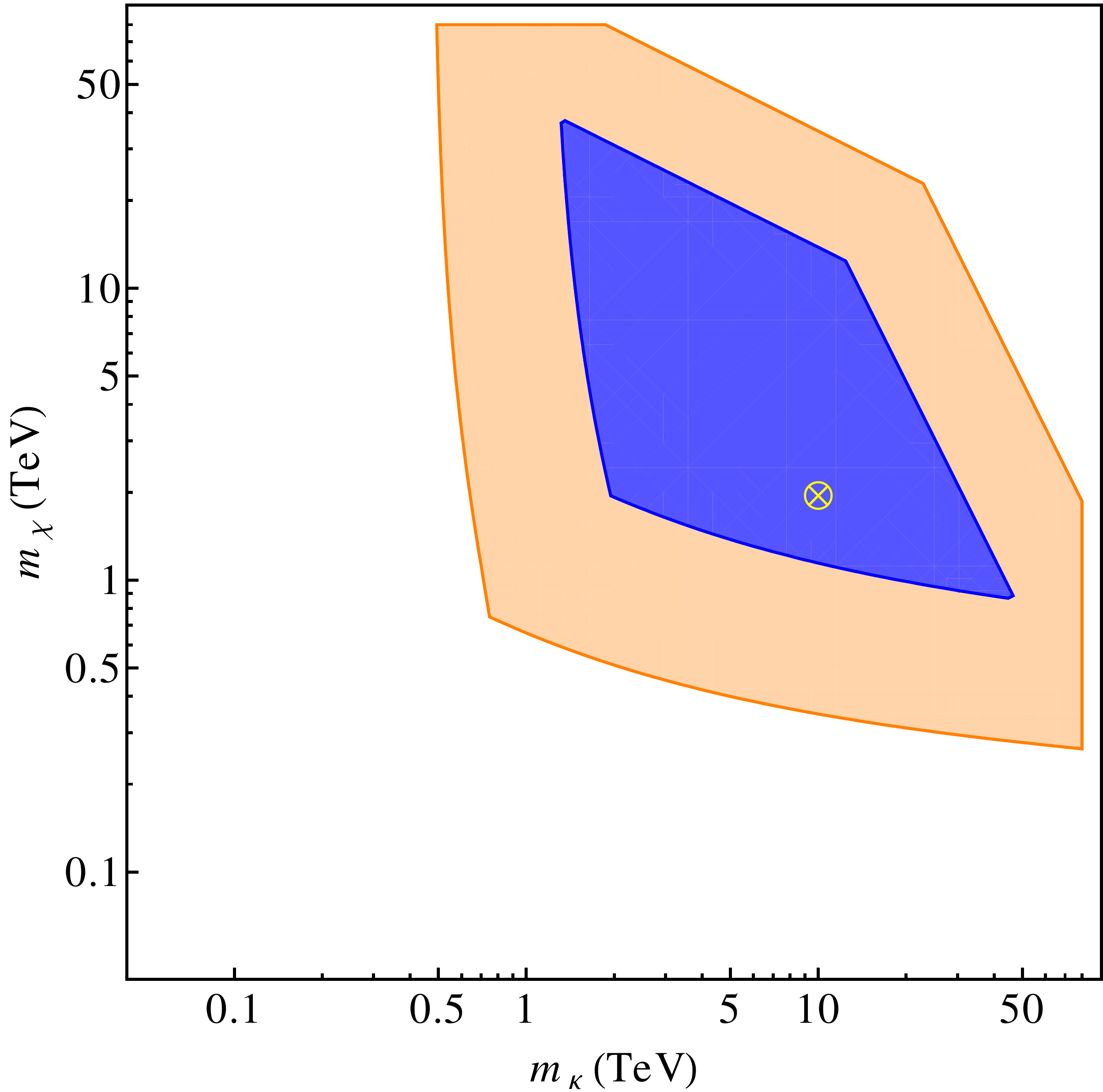}
\par\end{centering}

\caption{Limits on the masses of the two doubly-charged scalars of the model
(see text for details). \label{fig:doubly-charged-masses}}
\end{figure}

We can visualize the constraints on the scalar masses with the plot
in Figure~\ref{fig:doubly-charged-masses}, where we show the projection
on the $m_{\kappa}-m_{\chi}$ plane of the allowed parameter space
region by assuming perturbative unitarity and $\mu_{\kappa}<20$ TeV.
The blue, darker (orange, lighter) areas correspond to $v_{\chi}=2\ (5)\ {\mathrm{GeV}}$.
The cross stand for a reference point $m_{\kappa}=10\ {\mathrm{TeV}},m_{\chi}=2\ {\mathrm{TeV}}$
(and $v_{\chi}=2\ {\mathrm{GeV}}$, $\mu_{\kappa}=15\ {\mathrm{TeV}}$,
with $g_{ee}=1$ and $g_{e\mu}=0.001$).

\subsubsection{Constraints on the $\nu$ mass matrix}

The particular structure of the neutrino mass matrix $(m_{\nu})_{ab}\propto m_{a}g_{ab}^{*}m_{b}$
together with the limits on $g_{ab}$ coming from LFV processes almost
fix the structure of the neutrino mass matrix. Indeed, the requirement
of large $0\nu\beta\beta$ implies a relatively large $g_{ee}$ and
scalar masses relatively low, but on the other hand, $(m_{\nu})_{ee}$
is highly suppressed by the factor $m_{e}^{2}$, while $(m_{\nu})_{e\mu}$
is suppressed because so is $g_{e\mu}$ by the $\mu\rightarrow3e$
bound. Thus, taking into account all these limits the neutrino mass
matrix must fulfill 
\begin{equation}
|m_{\nu}|=\begin{pmatrix}<10^{-4} & <10^{-4} & \sim0.01\\
<10^{-4} & \sim0.01 & \sim0.01\\
\sim0.01 & \sim0.01 & \sim0.01
\end{pmatrix}\;\mathrm{eV}\,.\label{eq:final-nu-mass-matrix-model}
\end{equation}
What means that only the NH can be accommodated. Moreover, this structure
also provides two predictions: 
\begin{itemize}
\item One for the lightest neutrino mass, $m_{\mathrm{MIN}}=m_{1}\sim0.004\,\mathrm{eV}$,
and
\item Other for $\sin^{2}\theta_{13}$ and $\delta$, $\sin^{2}\theta_{13}\gtrsim0.008$.
\end{itemize}
In Figure~\ref{fig:s13-delta} we present the allowed $\sin^{2}\theta_{13}-\delta$
region for $|(m_{\nu})_{ee,e\mu}|=0$. The green,~darker region is
obtained when measured mixings and mass differences (except $\sin\theta_{13}$)
are varied within 1~$\sigma$; while the yellow,~lighter one is
obtained by varying them within 3~$\sigma$ (we use values from the
global fit performed in Ref.~\cite{Schwetz:2011zk}). For comparison,
we also draw the recent measurement of $\sin^{2}(\theta_{13})$ \textsc{\cite{Adamson:2011ig,Abe:2011sj,Abe:2011fz,An:2012eh,Ahn:2012nd}}
($\sin^{2}(\theta_{13})=0.023\pm0.003$, dashed lines). From the figure,
it seems that large values of $|\delta|$ are somehow favoured by
the model.

\begin{center}
\begin{figure}
\begin{centering}
\includegraphics[width=0.5\columnwidth]{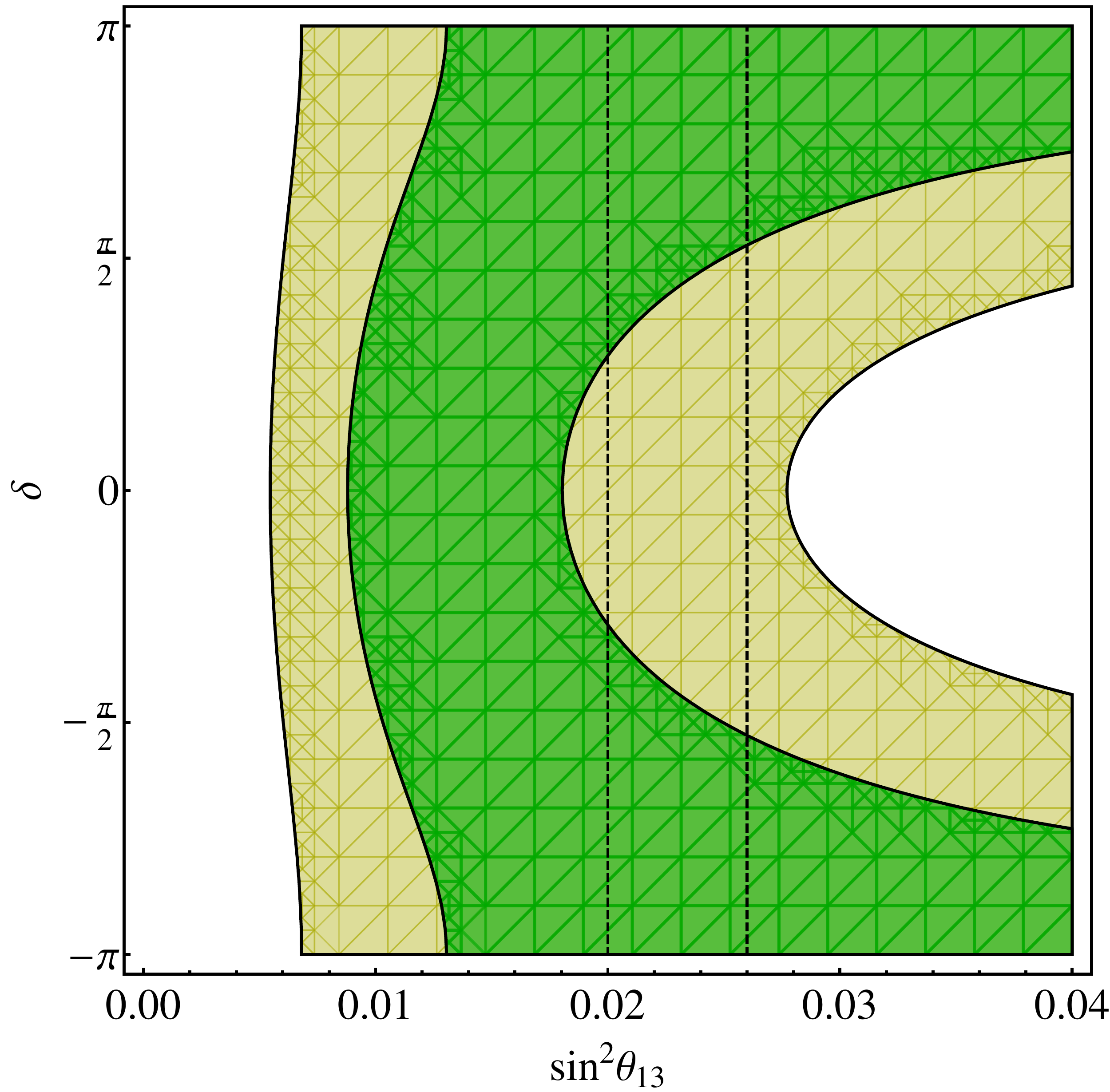}
\par\end{centering}

\caption{Allowed regions in the plane $\sin^{2}\theta_{13}-\delta$ when $|m_{ee}|,|m_{e\mu}|\ll0.01\,\mathrm{eV}$
(see text). \label{fig:s13-delta}}
\end{figure}

\par\end{center}

\section{Conclusions\label{sec:Conclusions}}

We have used the effective field theory language to classify NP contributions
to $0\nu\beta\beta$ involving operators without quarks. Charged lepton
chiralities, the operator dimension and the order at which $m_{\nu}$
should appear are linked for the lowest order operators:
\begin{itemize}
\item For $e_{L}e_{L}$: $0\nu\beta\beta$ appears at dimension 5 and $m_{\nu}$
at tree-level.
\item For $e_{L}e_{R}$: $0\nu\beta\beta$ appears at dimension 7, inducing
$m_{\nu}$ at one loop.
\item For $e_{R}e_{R}$: $0\nu\beta\beta$ appears at dimension 9, being
$m_{\nu}$ induced at two loops.
\end{itemize}
Hence, it is possible to have a sizable $0\nu\beta\beta$ while keeping
$m_{\nu}$ small (loop suppressed).The structure of the neutrino mass
matrix is in general constrained, and some of the parameters can be
predicted. Although often the models are complicated and tightly restricted,
they share a rich phenomenology affecting LFV processes and LHC searches,
especially if a doubly-charged scalar is discovered.

\section*{Acknowledgments}

This work has been supported in part by the Ministry of Economy and
Competitiveness, under the grant numbers FPA2006-05294, FPA2010-17915
and FPA2011-23897, by the Junta de Andalucía grants FQM 101 and FQM
6552, by the ``Generalitat Valenciana'' grant PROMETEO/2009/128,
and by the U.S. Department of Energy grant No.~DE-FG03-94ER40837.
A.A. is supported by the MICINN under the FPU program.

\providecommand{\href}[2]{#2}\begingroup\raggedright\endgroup

\end{document}